\documentclass[sigconf,natbib=true,anonymous=false]{acmart}
\makeatletter
\def\@ACM@checkaffil{
    \if@ACM@instpresent\else
    \ClassWarningNoLine{\@classname}{No institution present for an affiliation}%
    \fi
    \if@ACM@citypresent\else
    \ClassWarningNoLine{\@classname}{No city present for an affiliation}%
    \fi
    \if@ACM@countrypresent\else
        \ClassWarningNoLine{\@classname}{No country present for an affiliation}%
    \fi
}
\makeatother
\AtBeginDocument{%
  \providecommand\BibTeX{{%
    \normalfont B\kern-0.5em{\scshape i\kern-0.25em b}\kern-0.8em\TeX}}}

\copyrightyear{2024}
\acmYear{2024}
\setcopyright{rightsretained}
\acmConference[SIGIR '24]{Proceedings of the 47th International ACM SIGIR Conference on Research and Development in Information Retrieval}{July 14--18, 2024}{Washington, DC, USA}
\acmBooktitle{Proceedings of the 47th International ACM SIGIR Conference on Research and Development in Information Retrieval (SIGIR '24), July 14--18, 2024, Washington, DC, USA}
\acmDOI{10.1145/3626772.3657979}
\acmISBN{979-8-4007-0431-4/24/07}

\usepackage{latexsym}
\usepackage[skins]{tcolorbox}
\usepackage{balance}
\usepackage{graphicx}
\usepackage{natbib} 
\usepackage{spreadtab}
\usepackage{caption}
\usepackage[utf8]{inputenc}
\usepackage[T1]{fontenc}
\usepackage[english]{babel}
\usepackage{amsfonts}
\usepackage{nicefrac}
\usepackage{microtype}
\usepackage{amsmath}
\usepackage{mathtools}
\usepackage{subcaption}
\usepackage{booktabs}
\usepackage{ragged2e}
\usepackage{tikz}
\usepackage{stackengine}
\usepackage{etoolbox}
\usepackage{xspace}
\usepackage{xpatch}
\usepackage{cuted}
\usepackage{enumerate}
\usepackage{xstring}
\usepackage{setspace}
\usepackage{tabularx}
\usepackage{makecell}
\usepackage{changepage}
\usepackage{enumitem}
\usepackage{cuted}
\usepackage{cancel}
\usepackage{eqparbox}
\usepackage{bibentry}
\usepackage[hang,flushmargin]{footmisc}
\usepackage[capitalise,noabbrev,nameinlink]{cleveref}
\usepackage[useregional=numeric]{datetime2}
\usepackage[linesnumbered,ruled,noend]{algorithm2e}
\usepackage{adjustbox}
\usepackage{threeparttable}
\usepackage{booktabs, caption, makecell}
\usepackage{colortbl}

\usepackage[switch,mathlines]{lineno}
\usepackage{lipsum}
\usepackage{etoolbox}
\usepackage{longtable}

\graphicspath{{figures/}}

\makeatletter
\gdef\@copyrightpermission{
  \begin{minipage}{0.3\columnwidth}
   \href{https://creativecommons.org/licenses/by/4.0/}{\includegraphics[width=0.90\textwidth]{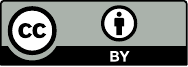}}
  \end{minipage}\hfill
  \begin{minipage}{0.7\columnwidth}
   \href{https://creativecommons.org/licenses/by/4.0/}{This work is licensed under a Creative Commons Attribution International 4.0 License.}
  \end{minipage}
  \vspace{5pt}
}
\makeatother

\begin{document}

\title[Can Query Expansion Improve Generalization of Strong Cross-Encoder Rankers?]{Can Query Expansion Improve Generalization \\of Strong Cross-Encoder Rankers?}


\author{Minghan Li}
\affiliation{%
  \institution{University of Waterloo}
}
\email{m692li@uwaterloo.ca}
\author{Honglei Zhuang}
\affiliation{%
  \institution{Google}
}
\email{hlz@google.com}
\author{Kai Hui}
\affiliation{%
  \institution{Google}
}
\email{kaihuibj@google.com}
\author{Zhen Qin}
\affiliation{%
  \institution{Google}
}
\email{zhenqin@google.com}
\author{Jimmy Lin}
\affiliation{%
  \institution{University of Waterloo}
}
\email{jimmylin@uwaterloo.ca}
\author{Rolf Jagerman}
\affiliation{%
  \institution{Google}
}
\email{jagerman@google.com}
\author{Xuanhui Wang}
\affiliation{%
  \institution{Google}
}
\email{xuanhui@google.com}
\author{Michael Bendersky}
\affiliation{%
  \institution{Google}
}
\email{bemike@google.com}
\settopmatter{authorsperrow=4}

\renewcommand{\shortauthors}{Minghan Li, Honglei Zhuang, Kai Hui, Zhen Qin, Jimmy Lin, Rolf Jagerman, Xuanhui Wang, Michael Bendersky}

\begin{abstract}
Query expansion has been widely used to improve the search results of first-stage retrievers, yet its influence on second-stage, cross-encoder rankers remains under-explored.
A recent work of~\citet{weller2023generative} shows that current expansion techniques benefit weaker models such as DPR and BM25 but harm stronger rankers such as MonoT5. In this paper, we re-examine this conclusion and raise the following question: Can query expansion improve generalization of strong cross-encoder rankers?
To answer this question, we first apply popular query expansion methods to state-of-the-art cross-encoder rankers and verify the deteriorated zero-shot performance.
We identify two vital steps for cross-encoders in the experiment: high-quality keyword generation and minimal-disruptive query modification.
We show that it is possible to improve the generalization of a strong neural ranker, by prompt engineering and aggregating the ranking results of each expanded query via fusion.
Specifically, we first call an instruction-following language model to generate keywords through a reasoning chain.
Leveraging self-consistency and reciprocal rank weighting, we further combine the ranking results of each expanded query dynamically.
Experiments on BEIR and TREC Deep Learning 2019/2020 show that the nDCG@10 scores of both MonoT5 and RankT5 following these steps are improved, which points out a direction for applying query expansion to strong cross-encoder rankers.
\end{abstract}

\vspace{-10mm}
\vspace{-10mm}
\ccsdesc[500]{Information systems~Retrieval models and ranking}
\vspace{-5mm}
\keywords{Query Expansion, Large Language Models, Cross-Encoder Rankers}
\vspace{-3mm}

\maketitle

\section{Introduction}\label{sec:intro}
Query expansion has been a core technique in information retrieval for over half a century~\cite{salton1971smart, robertson1976relevance, abdul2004umass}.
The goal is to increase the retrieval accuracy by adding additional terms to the query.
Conventional methods such as RM3~\cite{lavrenko2001rm3} leverage pseudo relevance feedback (PRF) to select terms from the documents retrieved for the original query as expansions~\citep{10.1145/3570724}.
Recently, large language models (LLMs) demonstrate their effectiveness in generating expansion terms for retrieval, which is known as generative query expansion~\citep{jagerman2023query,wang2023query2doc}.
However, both methods or their combinations are mainly considered for improving the recall and precision of \textit{first-stage retrievers}, yet their influence on the generalization of \textit{second-stage, cross-encoder rerankers} remains under-explored.

\begin{table}[!t]
\centering
\begin{adjustbox}{max width=0.43\textwidth}
\begin{tabular}{l|cc}
\toprule
\textbf{Methods}
&DL19 
&DL20 \\
\midrule
RankT5$_\text{base}$~\citep{zhuang2023rankt5}\qquad\qquad\qquad\qquad\qquad\qquad\qquad
&0.737
&0.736\\
\quad - w/ RM3~\citep{lavrenko2001rm3}
&0.681 
&0.695\\
\quad - w/ query2doc~\citep{wang2023query2doc}
&0.572 
&0.637\\
\quad - w/ query2keyword~\citep{jagerman2023query}
&0.690
&0.688\\
\quad - w/ Ours 
&\textbf{0.751} 
&\textbf{0.752}\\
\midrule
MonoT5$_\text{base}$~\citep{nogueira2020document} 
&0.695 
&0.720\\
\quad - w/ RM3 
&0.673 
&0.693\\
\quad - w/ query2doc
&0.473 
&0.613\\
\quad - w/ query2keyword 
&0.672 
&0.682\\
\quad - w/ Ours 
&\textbf{0.724} 
&\textbf{0.730}\\
\bottomrule
\end{tabular}
\end{adjustbox}
\caption{NDCG@10 on TREC DL 2019/2020. Directly applying existing query expansion methods on strong cross-encoder rankers can cause performance deterioration.}
\label{tbl:dl}
\vspace{-8mm}
\end{table}
This problem is interesting as additional terms usually contain more information about the query, yet it is rarely used in cross-encoder ranking. A recent study from \citet{weller2023generative} explores generative query expansion for different retrievers and rankers. They found that weaker models benefit more from expansions while stronger rankers are hurt in most times. This counter-intuitive observation calls for further examination of current expansion methods and whether there is a way to improve the results of strong rankers using query expansion.
To verify the conclusion, we also apply some popular query expansion methods to two state-of-the-art cross-encoder rankers, RankT5~\cite{zhuang2023rankt5} and MonoT5~\cite{nogueira2020document}. As shown in Table~\ref{tbl:dl}, the results are consistent with \citet{weller2023generative} where we observe that the nDCG@10 scores of both rankers on TREC DL 2019/2020 and BEIR are compromised using either PRF or LLMs. More results are shown in Section~\ref{sec:exp} and~\ref{sec:ablation}.

In this work, we identify two important steps for successfully using query expansion in cross-encoder rankers, which are not well explored by existing work focusing on retrievers: high-quality keyword generation and minimal-disruptive query modification.
First, cross-encoder rankers mainly aim to improve precision-related metrics such as nDCG@10 which are sensitive to noisy keywords, while retrievers care more about recall where some low-quality expanded keywords are less influential. Therefore, strong rankers such as RankT5 have higher demand on the generation quality.
Second, cross-encoder rankers are heavily based on token interactions, making them sensitive to the distributional shift in queries (e.g., number of tokens, input formats) compared to retrievers. Therefore, inserting documents in a query~\citep{wang2023query2doc} might be less desirable.

In our experiments, we find that the most effective way is to first use an LLM to generate high-quality, short keywords through a reasoning chain~\citep{wei2023chainofthought}. 
We then follow self-consistency~\citep{wang2023selfconsistency} to run the above process multiple times to filter noisy keywords and select top-k candidates, ensuring the quality of the generated keywords.
To mitigate distributional shift in query, we insert each keyword independently with the query and use reciprocal rank weighting~\citep{10.1145/1571941.1572114} to combine the ranking results.
Our pipeline manages to improve the nDCG@10 over the baselines for cross-encoder rankers such as RankT5~\cite{zhuang2023rankt5} on both BEIR and TREC DL 2019/2020, while other baselines that have been found effective for retrievers fail to improve such strong rankers.
Our study provides a preliminary yet novel research foundation for researchers to explore query expansion for cross-encoder rankers.

\section{Related Work}
\paragraph{Query Expansion and Fusion}
Early research on query expansion concentrated on utilizing either lexical knowledge bases~\citep{j.ipm.2006.09.003,6816606,10.5555/188490.188508} or Pseudo-Relevance Feedback (PRF)~\citep{imani2019deep,roy2016using,zheng2021contextualized}.
Recent studies show that scaling up LLMs through pre-training with more extensive and higher-quality corpora~\citep{anil2023palm,brown2020language,ouyang2022training,openai2023gpt4,touvron2023llama} can result in higher capabilities. Researchers have used large language models for generating keywords in the context of query expansion~\citep{wang2023query2doc,claveau2021neural,jagerman2023query,wang2023generative}.

The effectiveness of query variant fusion has been proven in previous works.
\citet{BELKIN1995431}  pioneered the fusion of multiple query variations into a single ranked list. 
\citet{10.1145/3077136.3080839} introduced Rank Biased Centroids for effective query variation fusion. 
\citet{10.1145/3166072.3166084} furthered this by applying reciprocal rank fusion~\citep{10.1145/1571941.1572114} and CombSUM~\citep{BELKIN1995431} with double fusion.

\paragraph{LLM-Based Neural Rankers}
MonoBERT \citep{nogueira2020passage} stands out as one of first cross-encoders for text reranking tasks. 
CEDR~\citep{MacAvaney_2019} introduces a more intricate approach by incorporating token representations at all layers of the Transformer using pre-BERT neural rerankers~\citep{Guo_2016}.
More potent rankers based on LLMs have emerged to directly score the relevance between queries and documents~\citep{zhuang2023rankt5,nogueira2020document,zhang2023improving}.
Most recently, LLMs have showcased remarkable efficacy when tasked with few/zero-shot text ranking such as LRL~\citep{ma2023zeroshot}, RankGPT~\citep{sun2023chatgpt}, RG-$k$L~\cite{zhuang2023beyond}, RankVicuna~\citep{pradeep2023rankvicuna}, and RankLlama~\citep{ma2023finetuning}. Alternatively, they can perform pairwise comparisons between passages, as demonstrated by PRP~\citep{qin2023large}. 
Despite the zero-shot effectiveness, the multiple decoding passes render them slow and non-parallelizable.

\section{Framework and Implementation}\label{sec:prelim}
\subsection{Cross-Encoder Ranking}
Given a query $q$, the text retrieval or ranking task is to return a sorted list of documents $\{d_1, d_2, . . ., d_{k}\}$ from a large text corpus $\mathcal{C}$ to maximize a metric of interest.
In this paper, we assume a set of candidate documents $\{d_1, d_2, . . ., d_{k}\}$ generated by a first-stage retriever are given and focus on the second-stage reranker to re-order the candidate documents.
The query-document pairs are encoded together for fine-grained token-level interactions:
\begin{align}\label{eq:sim_concat}
    s(q, d) = \phi (\text{concat}(q, d)),
\end{align}
where $\phi$ is the reranker and $s$ is the similarity score.
The ``concat'' function is implemented using special tokens as indicators, such as ``$\text{Question:}\ q \ \text{Document:}\ d$''.
In the following subsection, we will introduce the framework we use for keyword generation and selection to improve the results of cross-encoders. 

\subsection{High-Quality Keyword Generation}
The first step of query expansion is to generate keywords $\{w_1, w_2,...,$ $w_i\}$ semantically similar to the query $q$.
There are generally two sources of signals:
The classical approach which involves corpus-based signals through pseudo-relevance feedback (PRF) or
more recent approaches leveraging signals from LLMs by prompting~\cite{jagerman2023query}.
Notice that some LLM-based methods like Q2D~\citep{wang2023query2doc} and HyDE~\citep{gao-etal-2023-precise} generate much longer passages or documents rather than keywords, which drift too much from the distribution of queries and deteriorate cross-encoder reranker performance. In the following components of this framework, we do not consider expansion other than keywords, but will show the results of using excessively long expansion in Section~\ref{sec:exp}.
We explore four different methods for keyword generation, including both LLM-based and PRF-based methods, as well as their combinations:

\noindent$\bullet$ \textbf{PRF-based} methods like RM3~\citep{lavrenko2001rm3} to extract keywords from the retrieved documents $d_1, d_2,...$.\\
\noindent$\bullet$ \textbf{LLM-based} methods to generate keywords $w_1, w_2,...$  like Q2K~\citep{jagerman2023query}.\\
\noindent$\bullet$ \textbf{PRF + D2K}, which uses an LLM to extract keywords $w_1, w_2,...$ from the retrieved documents $d_1, d_2,...$. \\
\noindent$\bullet$ \textbf{Q2D2K}, which uses the LLM to generate detailed documents $d'_1, d'_2,...$ first and then selects a set of keywords $w_1, w_2,...$.

PRF + D2K and Q2D2K are inspired by Q2D~\citep{wang2023query2doc} and HyDE~\citep{gao-etal-2023-precise}. 
These methods are problematic when the generated documents are directly used as queries for cross-encoder rankers due to the query distribution shift, but are useful when the generated documents are summarized into keywords for expansions.
The assumption is that documents generated by a well pre-trained LLM can already answer the question or at least contain helpful keywords.
The intuition is also similar to recitation-augmented language models~\citep{sun2023recitationaugmented} where more knowledge can be elicited before fulfilling the task.

\begingroup
\setlength{\tabcolsep}{2pt} 

\begin{table*}[!t]
\centering
\begin{adjustbox}{max width=0.97\textwidth}
\begin{tabular}{@{}l|ccccccccccccccccc|c@{}c@{}}
\toprule
\textbf{Methods}
& AA 
& SF 
& NQ 
& Fe 
& Qu 
& HQ 
& FQ 
& CF 
&BA 
& SD 
& T2 
&TN 
&S1 
& DB 
& NF 
&R0 
&TC 
&BEIR 
&Wiki+News\\
\midrule
RankT5$_\text{base}$
&0.323
&0.750
&0.565
&0.826
&0.814
&0.732
&\textbf{0.414}
&0.255
&0.541
&0.174
&0.383
&0.439
&\textbf{0.299}
&0.448
&0.375
&0.525
&\textbf{0.782}
&0.508
&0.557\\
\quad - w/ RM3
&\textbf{0.330}
&0.722
&0.546
&0.817
&\textbf{0.831}
&0.689
&0.377
&0.257
&0.473
&0.165
&0.360
&0.443
&0.291
&0.393
&0.358
&0.481
&0.746
&0.487
&0.539\\
\quad - w/ query2doc
&0.271
&0.671
&0.511
&0.810
&0.749
&0.640
&0.300
&0.184
&0.358
&0.140
&0.275
&0.419
&0.231
&0.367
&0.331
&0.447
&0.716
&0.437
&0.502\\
\quad - w/ query2keyword
&0.315
&0.739
&0.546
&0.841
&0.807
&0.725
&0.381
&0.257
&0.500
&0.168
&0.353
&\textbf{0.470}
&0.280
&0.398
&0.365
&0.496
&0.741
&0.493
&0.556\\
\quad - w/ Ours 
&0.324
&\textbf{0.752}
&\textbf{0.577}
&\textbf{0.846}
&0.822
&\textbf{0.744}
&0.412
&\textbf{0.261}
&\textbf{0.542}
&\textbf{0.176}
&\textbf{0.390}
&0.454
&0.292
&\textbf{0.452}
&\textbf{0.377}
&\textbf{0.541}
&0.781
&\textbf{0.514}*
&\textbf{0.570}*\\
\midrule
MonoT5$_\text{base}$
&0.242
&0.713
&0.553
&0.808
&0.785
&0.705
&0.364
&0.216
&0.495
&0.161
&\textbf{0.406}
&0.417
&0.286
&0.423
&0.312
&0.415
&\textbf{0.685}
&0.470
&0.519\\
\quad - w/ RM3 
&\textbf{0.249}
&0.698
&0.544
&0.817
&\textbf{0.799}
&0.666
&0.354
&0.219
&0.453
&0.156
&0.390
&0.427
&\textbf{0.291}
&0.412
&\textbf{0.314}
&0.396
&0.632
&0.465
&0.511 \\
\quad - w/ query2doc
&0.188
&0.640
&0.508
&0.784
&0.353
&0.594
&0.302
&0.108
&0.353
&0.127
&0.339
&0.398
&0.242
&0.368
&0.293
&0.391
&0.590
&0.392
&0.459\\
\quad - w/ query2keyword 
&0.232
&0.697
&0.542
&0.828
&0.774
&0.702
&0.341
&0.215
&0.449
&0.156
&0.378
&\textbf{0.453}
&0.284
&0.397
&0.293
&0.359
&0.642
&0.455
&0.517\\
\quad - w/ Ours 
&0.240
&\textbf{0.720}
&\textbf{0.569}
&\textbf{0.831}
&0.793
&\textbf{0.719}
&\textbf{0.369}
&\textbf{0.223}
&\textbf{0.499}
&\textbf{0.163}
&0.404
&0.431
&0.288
&\textbf{0.443}
&0.313
&\textbf{0.420}
&0.671
&\textbf{0.476}*
&\textbf{0.532}*\\
\bottomrule
\end{tabular}
\end{adjustbox}
\caption{NDCG@10 on BEIR.
TC=TREC-COVID, 
NF=NFCorpus, 
NQ=NaturalQuestions, 
HQ=HotpotQA, 
FQ=FiQA, 
AA=ArguAna, 
T2=Touché-2020, 
Qu=Quora, 
DB=DBPedia, 
SD=SCIDOCS, 
Fe=FEVER, 
CF=Climate-FEVER, 
SF=SciFact, 
S1=Signal-1M, 
BA=BioASQ, 
R0=Robust04, 
TN=TREC-NEWS.
$*$: pass the paired t-test against the other baselines ($p<0.01$).}
\label{tbl:main}
\vspace{-5mm}
\end{table*}
\endgroup

As mentioned in Section~\ref{sec:intro}, the precision-related metrics, which cross-encoder rankers aim to optimize, are more sensitive to the noise in the expanded keywords.
If there are noisy keywords generated from the previous stage, the performance of rankers are more likely to be affected compared to retrievers. 
Hence, we argue that it is necessary to add a filtering stage in this framework to remove noisy keywords and increase the reliability of the expansion.
We leverage self-consistency~\citep{wang2023selfconsistency} in LLM literature for filtering. 
For LLM-based keyword generation methods which involves stochasticity, we repeat the keyword generation method multiple times and select the top-$k$ keywords that have the highest majority votes (i.e., frequency).
For deterministic methods like RM3, we simply take the keywords with the highest RM3 keyword weights.

\subsection{Minimal-Disruptive Query Modification}
One way to insert keywords is to directly concatenate them with the query. 
The concatenation function we use is ``$\text{Question:}\ q \ w_1 \ w_2 \ ... w_i$ $ \ \text{Document:}\ d$''.
However, as mentioned in Section~\ref{sec:intro}, the increasing number of keywords or excessively long expansion may overwhelm the original query, especially for cross-encoder models which rely more on query-document token-interaction.
In Section~\ref{sec:exp}, we will show that even increasing the number of keywords to 3 will result in degraded precision.

To mitigate the distributional shift in query, another way is to concatenate each keyword individually with the query and fuse the final ranking results together~\citep{10.1145/3341981.3344224}. Inserting only 1 keyword is the minimal disruptive expansion we found for cross-encoder, which is proved to be very robust on multiple datasets. Specifically, the new similarity scoring function will be:
\begin{align}\label{eq:sim_fusion}
    s(q, d) =  \sum_i \alpha_i\cdot\phi (\text{concat}(q, w_i, d)),
\end{align}
where the concatenation function $\text{concat}(q, w_i, d)$ is implemented as ``$\text{Question\\ :}\ q\ w_i \ \text{Document:}\ d$'' and $\alpha_i$ is the weight for the expansion $w_i$.
We find this formulation more effective than concatenating all keywords at once as the number of keywords increases.

Previous works~\citep{zhuang2021ensemble,bendersky2020rrf102} have also found that ensembling runs from different models or data augmentation can be effective for ranking.
After obtaining the candidate keywords, we concatenate each keyword independently with the original query before ranking. 
For each expanded query, we first use it in the cross-encoder ranker model to rerank the top-1000 candidate documents retrieved by BM25 to get a ranked list. 
For the fusion weights in Equation~\eqref{eq:sim_fusion}, we follow the previous work~\citep{10.1145/1571941.1572114,dai2022promptagator} to weight the ranked lists using the reciprocal rank of the top-1 document in retrieved list:
\begin{align}\label{eq:rr}
    \alpha_i = \frac{1}{\text{Rank}(d^+, D_i)}
\end{align}
where Rank$(d^+, D_i)$ is the rank of the top-1 document $d^+$ retrieved for the original query in a candidate list of expansion $w_i$.
Finally, we combine the ranking list of the original query in case all the expansions are not helpful.

\section{Experiments}\label{sec:exp}
\paragraph{Models and Datasets} For keyword generation, we use Flan-PaLM2-S~\citep{anil2023palm}.
For cross-encoder ranking, we test two different rankers: MonoT5~\citep{nogueira2020document} and RankT5~\citep{zhuang2023rankt5}.
We reproduce the MonoT5 model using the point-wise loss in~\citet{zhuang2023rankt5}.
For in-domain evaluation, we evaluate on TREC DL 2019 and 2020~\citep{craswell2020overview}, containing 43 and 54 test queries, respectively. The relevance sets are densely labelled with scores from 0 to 4.
For out-of-domain evaluation, we evaluate on 17 datasets from BEIR.

\begin{table}[!t]
\centering
\begin{adjustbox}{max width=0.43\textwidth}
\begin{tabular}{l|cc|cc}
\toprule
Methods &DL19 &DL20 & BEIR & Wiki+News\\\midrule
Q2D2K-fusion &\textbf{0.751} &\textbf{0.752}&\textbf{0.514}&\textbf{0.570}\\
Q2K-fusion &0.750 &0.748 &0.510 &0.5628\\
PRF + D2K-fusion &0.745 &0.733 &0.510 &0.565\\
RM3-fusion &0.741 &0.737 &0.510 &0.560\\
\bottomrule
\end{tabular}
\end{adjustbox}
\caption{NDCG@10 score of Q2D2K, Q2K, RM3, and PRF + D2K with fusion based on RankT5.}
\label{tbl:rq1}
\vspace{-8mm}
\end{table}

\paragraph{Evaluation} We report the nDCG@10 metric~\cite{jarvelin2002cumulated} as many datasets in BEIR and TREC DL are densely labeled, and the top-10 setting reflects the common use case in applications.
As the LLM is fine-tuned on instructions set which has large overlap with the Wikipedia and News corpus in BEIR, we also report the average score on Natural Questions, FEVER, Climate-FEVER, HotpotQA, TREC-News, and Robust04 datasets (Wiki+News) besides the main score.

\paragraph{Inference Pipeline}\label{sec:pipeline}
We use BM25's top-1000 retrieval results as the candidates for reranking using Pyserini~\citep{10.1145/3404835.3463238}.
For RM3, we use Pyterrier~\citep{10.1145/3459637.3482013} to extract keywords from the retrieved documents.

For keywords generation method Q2D2K, we follow the instruction template in Promptagator~\citep{dai2022promptagator} and ask Flan-PaLM2-S to generate 2 documents based on the query and then extract 5 keywords for each document. We then repeat the process 3 times to obtain 30 keywords.
For keyword generation method PRF + D2K, we replace the generated documents by BM25 retrieved documents.
By default, we use Q2D2K as the keyword generation component.
We then run self-consistency and select the top-3 keywords from the keyword candidates and concatenate each of them individually with the query before feeding them into the ranker.

For fusion, we use the reciprocal ranks of the top-1 document retrieved for the original query in each expansion's reranked list as weights.
We finally fuse the aggregated expansion results with the original reranking list from the cross-encoder ranker as regularization, with a coefficient of 0.3. 
Coefficients range from 0.1 to 0.3 work similarly but not included due to space limit.

\paragraph{Results} Table~\ref{tbl:dl} and~\ref{tbl:main} show the main results on TREC DL19/20 and BEIR benchmark, where directly applying the query expansion pipeline in retrieval to cross-encoder ranker reranking results in deteriorated performance even with top-3 keyword concatenation, while our method can improve over the original cross-encoder ranker scores on both TREC DL and BEIR.
The improvement on TREC DL and Wiki+News BEIR datasets are more significant compared to other datasets in BEIR, which results from the instruction fine-tuning step of PaLM2 as previously mentioned.

\section{Ablation Analysis}\label{sec:ablation}

\paragraph{RQ1: Whether to use LLMs or PRF for keyword generations?}
To compare the keyword generation quality of different methods, we fix the keyword insertion and fusion procedure while varying the keyword generation methods. Table~\ref{tbl:rq1} shows the comparison results.
We can see that the keywords generation quality is reflected in the nDCG@10 scores, where 
our Q2D2K method manages to outperform other LLM or PRF based methods on DL 19/20 and BEIR, reflecting higher keyword quality.
Besides that, fusion is also very important for maintaining cross-encoder ranker zero-shot performance which will be discussed in detail in RQ2.

\begin{figure}[t!]
\centering
\hspace{-0.4cm} 
\includegraphics[width=.43\textwidth]{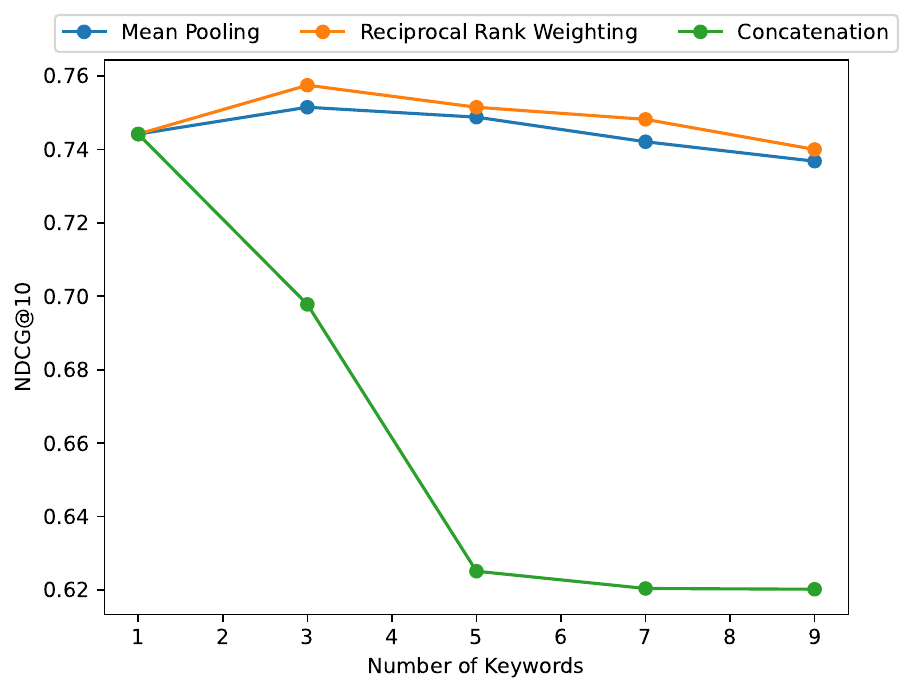}
\vspace{-4mm}
\caption{NDCG@10 score on DL19 using different number of keywords for RankT5.}
\label{fig:dl19}
\vspace{-6mm}
\end{figure}

\paragraph{RQ2: How to use these keyword expansions?}
Figure~\ref{fig:dl19} plots the number of keywords used for ranking fusion and its influence on the final nDCG@10 score.
We can see that the improvement of our proposed method (which uses Reciprocal Rank Weighting) peaks at 3 expanded keywords and gradually diminishes with the addition of more keywords.
Although the generated keywords are useful, they still brings noise which escalates as the keyword number increases.

\begin{table}[!t]
\centering
\begin{adjustbox}{max width=0.43\textwidth}
\begin{tabular}{l|cc|cc}
\toprule
Methods &DL19 &DL20 & BEIR & Wiki+News\\\midrule
RankT5$_\text{base}$ &0.737 &0.736 &0.508&0.557\\
\quad +mean pooling &0.748 &0.750 &0.514
&0.572\\
\quad +reciprocal rank &0.751 &0.751 &0.514 &\textbf{0.573} \\
\quad +original query &\textbf{0.751} &\textbf{0.752} &\textbf{0.514} &0.570 \\
\midrule
MonoT5$_\text{base}$ &0.695 &0.720&0.470&0.519\\
\quad +mean pooling &0.700 &0.727 &0.471&\textbf{0.533}\\
\quad +reciprocal rank &0.713 &\textbf{0.733} &0.475&0.532\\
\quad +original query &\textbf{0.724} &0.730 &\textbf{0.476}
&0.532 \\
\bottomrule
\end{tabular}
\end{adjustbox}
\caption{Incremental ablation on the keyword fusion process.
Details are introduced in Section~\ref{sec:ablation}.}
\label{tbl:rq2}
\vspace{-9mm}
\end{table}
\begin{table}[!t]
\centering
\begin{adjustbox}{max width=0.49\textwidth}
\begin{tabular}{l|ll}
\toprule
Methods &DL19 &DL20\\\midrule
Mean pooling &0.748 &0.750\\
Reciprocal rank\qquad\qquad\qquad\qquad\qquad &\textbf{0.751}\qquad\qquad 
&\textbf{0.751} \\
\midrule
Top-k overlap &0.741 &0.750\\
Ranker's entropy &0.729 &0.746 \\
KL divergence &0.741 &0.749 \\
WS distance &0.743 &0.751 \\
\bottomrule
\end{tabular}
\end{adjustbox}
\caption{Weighting methods for fusion on DL 19/20 based on RankT5. NDCG@10 scores are reported.
}
\label{tbl:fusion}
\vspace{-5mm}
\end{table}
Table~\ref{tbl:rq2} shows the ablation of the keyword fusion process, which mainly includes ranking average, reciprocal rank weighting, and combining with original query ranking results.
Adding mean fusion and reciprocal weighting consistently brings improvement to the model.
As for the original query fusion, we view it as an expansion and combine its ranking with the other expansion's ranking results using a default value of 0.3 for zero-shot evaluation. 
The improvement is less consistent as we did not perform hyper-parameter search on the fusion coefficient due to zero-shot evaluation, 
but instead use a default value of 0.3 for combining with the original query ranking results.
Figure~\ref{fig:dl19} also shows that fusion is more robust compared to keyword concatenation as the keywords increase.

Table~\ref{tbl:fusion} shows different fusion methods we tried on TREC DL 19/20. Mean pooling and reciprocal rank weighting are reported in Table~\ref{tbl:rq2}. For top-k overlap, we take the overlap between the original query's candidate list and the other expansion's candidate lists as weights for $\alpha_i$ in Equation~\eqref{eq:sim_fusion}. For ranker's entropy, we normalized the retrieved scores into a distribution for each expansion and use the reciprocal entropy of the distribution as weights.
For KL divergence and Wasserstein distance, they are similar to the ranker's entropy except that they calculate the distances between the original query's distribution and the other expansion's distributions. We also take the reciporal of this distance as weights for fusion.
We can see that among all the fusion techniques, the reciprocal rank weighting method has the best nDCG@10 scores on both DL 19 and 20, demonstrating the robustness of this simple fusion method.

\section{Conclusion}
In this paper, we examine the possibility of improving the generalization of cross-encoder rankers using query expansion based on the study of~\citet{weller2023generative}.
Our solution is to leverage an LLM to generate high-quality, concise keywords through a reasoning chain and individually evaluate the ranking scores of each expansion before aggregating them together.
We observe significant improvement on BEIR and TREC DL 2019/2020 over directly using the popular query expansion methods.


\bibliographystyle{ACM-Reference-Format}
\bibliography{sample-base}

\end{document}